# THE VARIATION OF THE SOLAR DIAMETER AND IRRADIANCE: ECLIPSE OBSERVATION OF JULY 11, 2010

SERGE KOUTCHMY

*Institut d'Astrophysique de Paris, UMR 7095, CNRS and UPMC, 98 Bis Bd Arago, F75014 Paris (France)*

CYRIL BAZIN, JEAN-YVES PRADO, PHILIPPE LAMY, PATRICK ROCHER

*Consortium for the observation of the July 11, 2011 Solar total eclipse,*
*Paris IAP and IMCCE (France), Toulouse- CNES(France), Marseille LAM (France),*

The variation of the solar diameter is the subject of hot debates due to the possible effect on the Earth climate and also due to different interpretations of long period solar variabilities, including the total irradiance. We shortly review the topic and show that rather long term variations, corresponding to a length well over a solar magnetic cycle, are interesting to consider. The very recently launched mission "Picard" is entirely devoted to the topic but will just permit a short term evaluation.

At the time of the last solar total eclipse of 11/7/2010, several experiments were prepared to precisely measure the transit time of the Moon related to the precise value of the solar diameter. Preliminary results coming from the use of a specially designed CNES photometer, put on different atolls of the French Polynesia, are presented. In addition the results of new experiments devoted to fast observations of flash spectra, including their precise chrono-dating, are illustrated and discussed. A new definition of the edge of the Sun, free of spurious scattered light effects strongly affecting all out of eclipse evaluations, is emerging from these observations, in agreement with the most advanced attempts of modeling the outer layers of the photosphere. We also argue for a definite answer concerning the solar diameter measurement from eclipses based on a better precision of lunar profiles coming from lunar altimetry space experiments which will be possible in the following decades.

## 1. Introduction

### 1.1. *Some historical background*

During the last decades, considerable efforts have been devoted to the analysis of possible effects of the rather irregular activity cycles of the Sun, through the solar forcing effects. They are related to the question of the anthropogenic origin of global warming, its amplitude and its consequences for the climate and economic life. These seemingly legitimate studies have an intriguing historical background that can be considered as premises for the still disputed methods of evaluation used in these studies, including the influence of the solar variabilities that we will discuss in this paper.

In this context, it is difficult to avoid a critical interpretation of the famous episode of the reign of the Pharaoh Akhenaton (Amenophis IV) in 1350 B.C. as a dramatic example of where an excessive belief can lead. In the ancient Egypt of Akhenaton the Sun has been imposed as a single divinity, see Figure 1, leading to a catastrophic end of the reign of this famous Pharaoh, although considered as the 1st historical attempt to impose a more rational pre- monotheistic society, well before the Christian era.

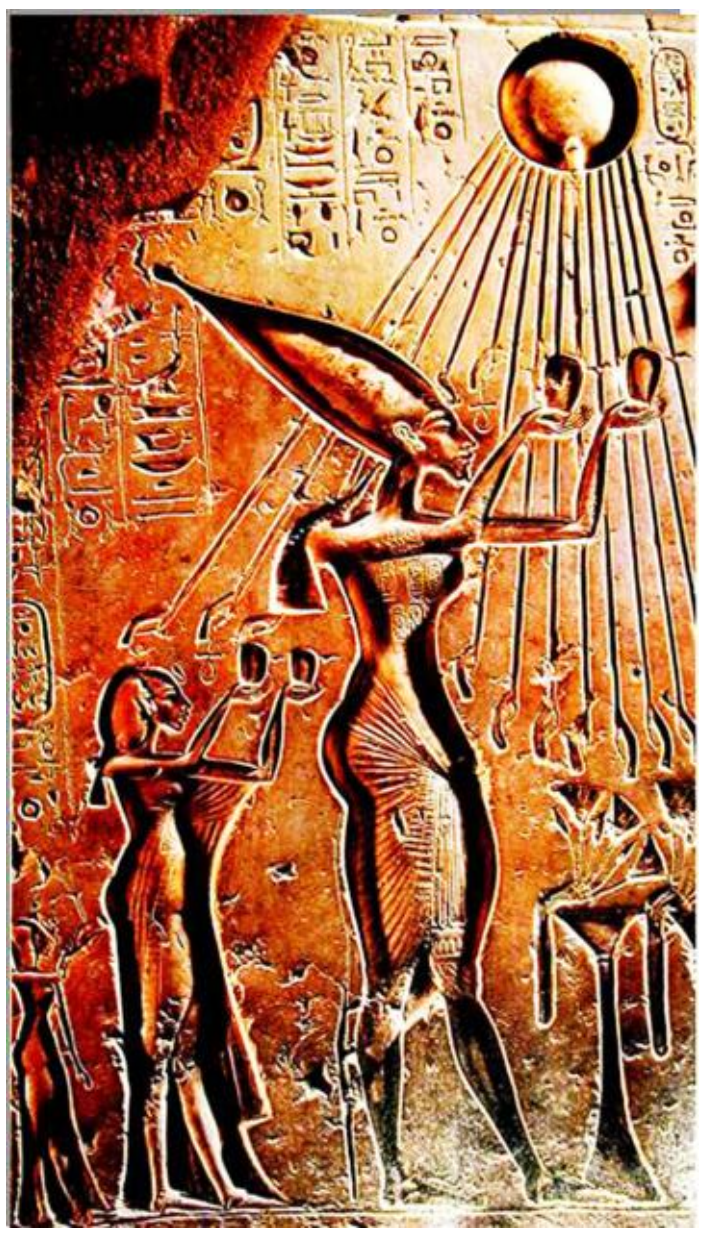

Figure 1- Akhenaton and family 1350 B.C. (Cairo Museum)

Furthermore, following the teaching of famous antique philosophers like Plato, Aristotle and their followers, the Man and not the Sun has been put at the center of the Word as yet reflected in modern religions (Christianism; Islam…). It roughly means that phenomena influencing our life should rather find their origin and result in the human activity. Unfortunately, the excessive belief in geo-centrism (a cosmogonic model imposed during almost 1.5 millennium) leads to wrong conclusions about the real World. The great discoveries of the 16$^{th}$ century changed the situation. Philosophy and Science slowly became disconnected and more progress has been made possible until today.

### 1.2. *The global warming epoch*

As regards to the global warming, it is now widely accepted (but not really unanimously so) that a global irreversible effect, see Fig. 2 will "soon" result as a consequence of the well confirmed rising amount of $CO_2$ (and other minor components of the Earth atmosphere like $NO^+$).

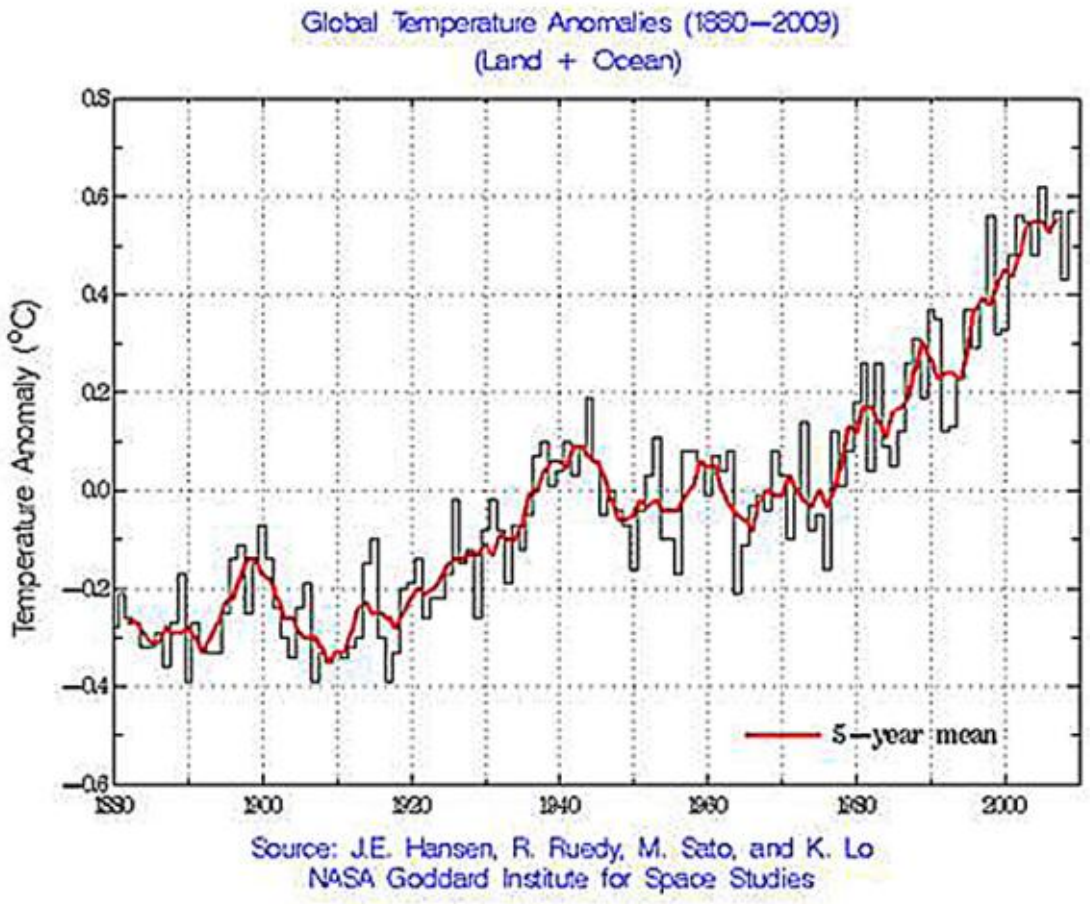

Figure 2- The global warming effect illustrated by the risisng average temperature recorded on the Earth.

These gazes produce an increasing greenhouse effect in the Earth atmosphere. $CO_2$ see Fig.3, is mainly produced by the industrial and technological activities based on the use of fossil origin energies (coal, oil, natural gaz, etc.). This is in contrast with the so-called durable and renewable energies (wind and solar energies, both thermal and photo-voltaic, geo-thermal energy etc.) that do not produce any greenhouse effect gaz.

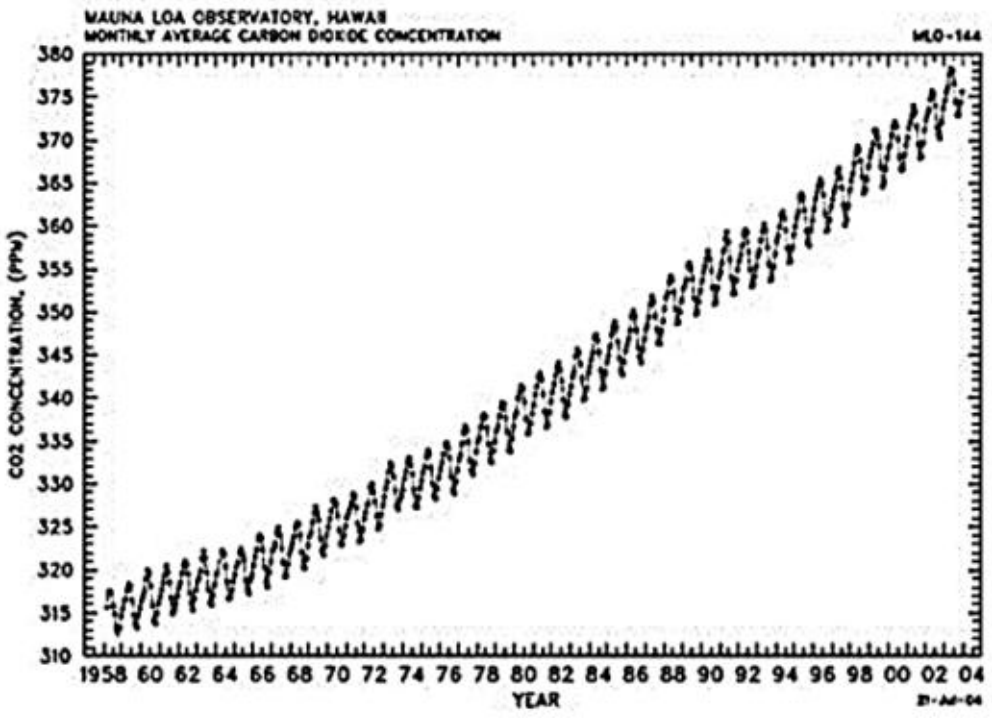

Figure 3- The recorded at Mauna Loa (NOAA) increasing amount of $CO_2$. Note the Yearly modulation superposed to the long range increase.

Climato-sceptics however exist and a seemingly reasonable opinion is possibly reflected by the following statement coming from the Policy Report No 321 by Iain Murray and H. Sterling Burnett (USA) from the "National Center for Policy Analysis", 2009:

"Global warming is a reality. But whether emissions of carbon dioxide $CO_2$ and other greenhouse gazes from human fossil fuel use are the principal cause- are uncertain." However even if it is uncertain it is a source of big concern!

### 1.3. *About the solar forcing and the activity cycles.*

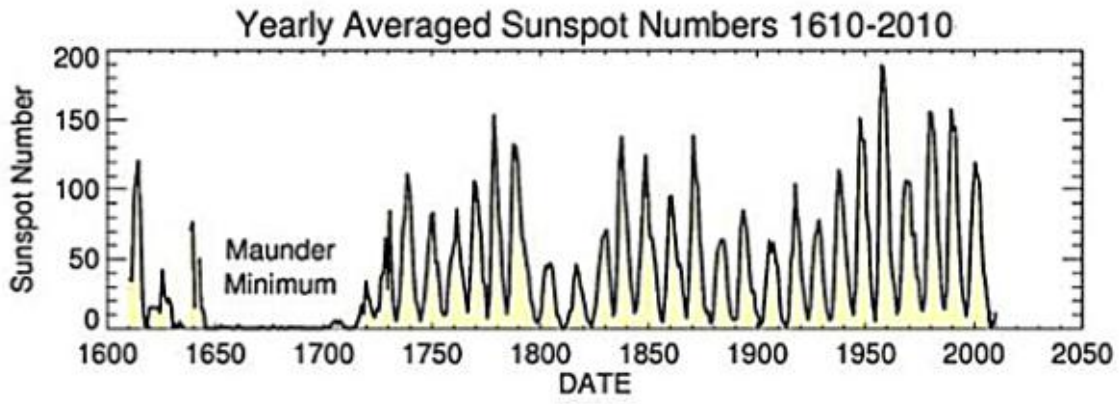

Figure 4- Sunspot cycles as shown using the sunspot numbers. The Maunder minimum epoch is well shown. See also *http://www.sidc.oma.be/*

Indeed there are data suggesting a possible solar effect to explain the long term variations of the climate on the Earth. The most popular example is coming from historical reports showing that an extended period of global cooling was recorded during almost 70 Years in the Western Word. It is also sometimes called the “little Ice Age”, and it coincides with the time of the so-called Maunder minimum [1] see Fig. 4, reflected by the well established series of measurements of the co-called international sunspot number now kept at the Brussels Royal Observatory and also “produced” by several US Agencies. This number usually shows a pronounced 11.3 Years periodicity due to a global dynamo effect producing the solar magnetic activity cycle with the same periodicity. To illustrate this climate abnormal cooling in the North Atlantic regions a famous picture of the frozen Thames by Jan Grifier is often shown and even more details of the sunspot cycle are tentatively interpreted [1], like a longer pseudo- periodicity near 100 Years. It is possible that this episode was over interpreted although some more convincing studies are now supporting the assumption of a significant solar forcing. It is first the dendrochronological data-tree rings results concerning the cosmogonic abundance of C14 extended over a much longer period of time, see Fig. 5.

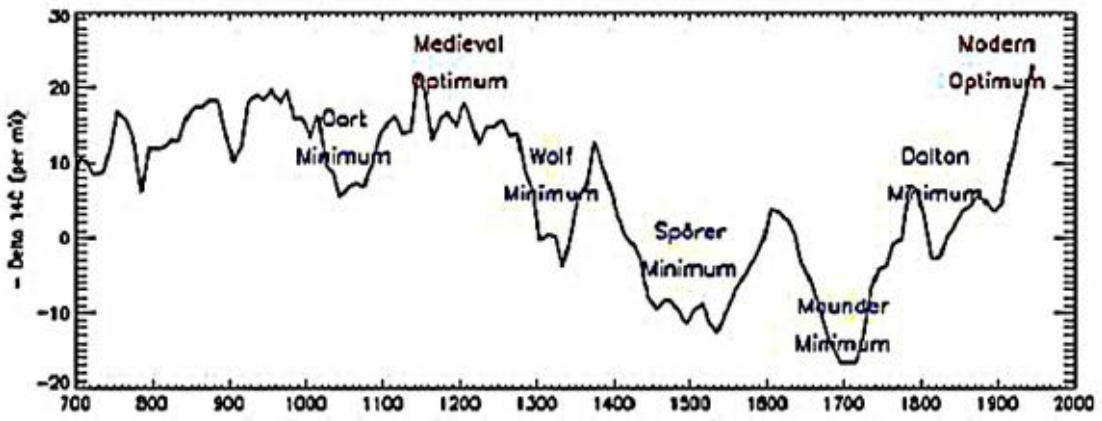

Figure 5- Ten Years average content of the C14 isotope deduced from the study of tree rings (from Intcal98, Quaterly Isotope Lab.).

Note in particular the time of the Medieval Optimum of warming in the 11- 12$^{th}$ centuries when the famous Vikings populated the Greenland and Iceland and developed farming. Also, as recorded in books from merchants, wines were produced in England, etc. No doubt that an extended period of relative warming existed at least in the Northern parts of the Atlantic Ocean Regions. A more extended study would consider several millennia and this has indeed been done using the analysis of proxies of the solar activity such as the amount of cosmogenic isotopes C14 and Be10 in natural stratified archives (e.g. tree rings and ice cores) [2] to discuss the influence of the solar activity during the Holocene. We will not go so far as new methods and techniques permit today to re-consider some more trivial aspects of this influence starting with the discussion of the irradiance variabilities, their origin and consequences.

## 2. Irradiance variations and heliometry.

The most obvious influence that the solar variabilities can have on the Earth climate is related to the slightly variable amount of thermal energy falling on the Earth.

This energy is measured using the so-called Total Solar Irradiance (TSI) measurements leading to the popular “solar constant” determination when the total solar spectrum is taken into account and summed. Note that the Yearly variation should first be removed when discussing the topic. It is easy to make it because this rather large modulation (of order of 7%) is very well reproducible and it depends only of the well established with a high precision orbital motion of the Earth around the Sun. All irradiance measurements are corrected for this geometrical effect and only variations extended well over 1 Year are discussed. An excellent former and rather introductory review to the topic can be found in [3]. The book suggested that the Sun has a changing diameter of large magnitude over the centuries and it has been the basis for proposing a space mission called “Picard” (the name of the 1$^{st}$ scientist who performed at the Paris Observatory a long series of measurements of the diameter of the Sun) to the French Space Agency CNES in order to perform solar diameter measurements free of Earth atmospheric effects (distortions, refraction, image motion due to the turbulence, smearing, scattering etc.). The topic was recently reviewed again e.g. [4]. The dedicated space mission Picard was launched in June 2010. Fortunately, solar eclipses offer another opportunity to perform these measurements free of Earth atmospheric effects. Here we now try to discuss the most recent aspects of the question.

### *2.1- About the possible variations of the solar diameter.*

Some correlated variations of the solar total irradiance S and of the solar diameter (or equivalently, the corresponding solar radius $R_0$) are suggested by a 1st approximation analysis of the variation of the total flux arriving at the Earth orbit or a distance of 1 astronomical units A. Assume the Sun is a perfect spherical constantly emitting body (neglecting the centre limb effect) with a temperature $T_{eff}$ of the photosphere, we use the Stefan- Boltzman constant σ to express the irradiance:

$$S = \pi R_0^2 \sigma T_{eff}^4 / A^2 \qquad (1)$$

A is constant and the relative variations of the irradiance $\delta S/S$ can then be due to variations of $T_{eff}$ or to $R_0$ or to both:

$$\delta S/S = 2\, \delta R_0/R_0 + 4\, \delta T_{eff} / T_{eff} \qquad (2)$$

The most recent measured TSI variations see Fig. 6, shows a typical 0.1 % maximum amplitude over a full solar sunspot cycle. Assume no variation of $T_{eff}$ we can immediately see from (2) that the maximum possible variation $\delta R_0$ of the solar radius cannot be more than 0"4 (the average radius of the Sun is approximately 960") and indeed it is considerably less

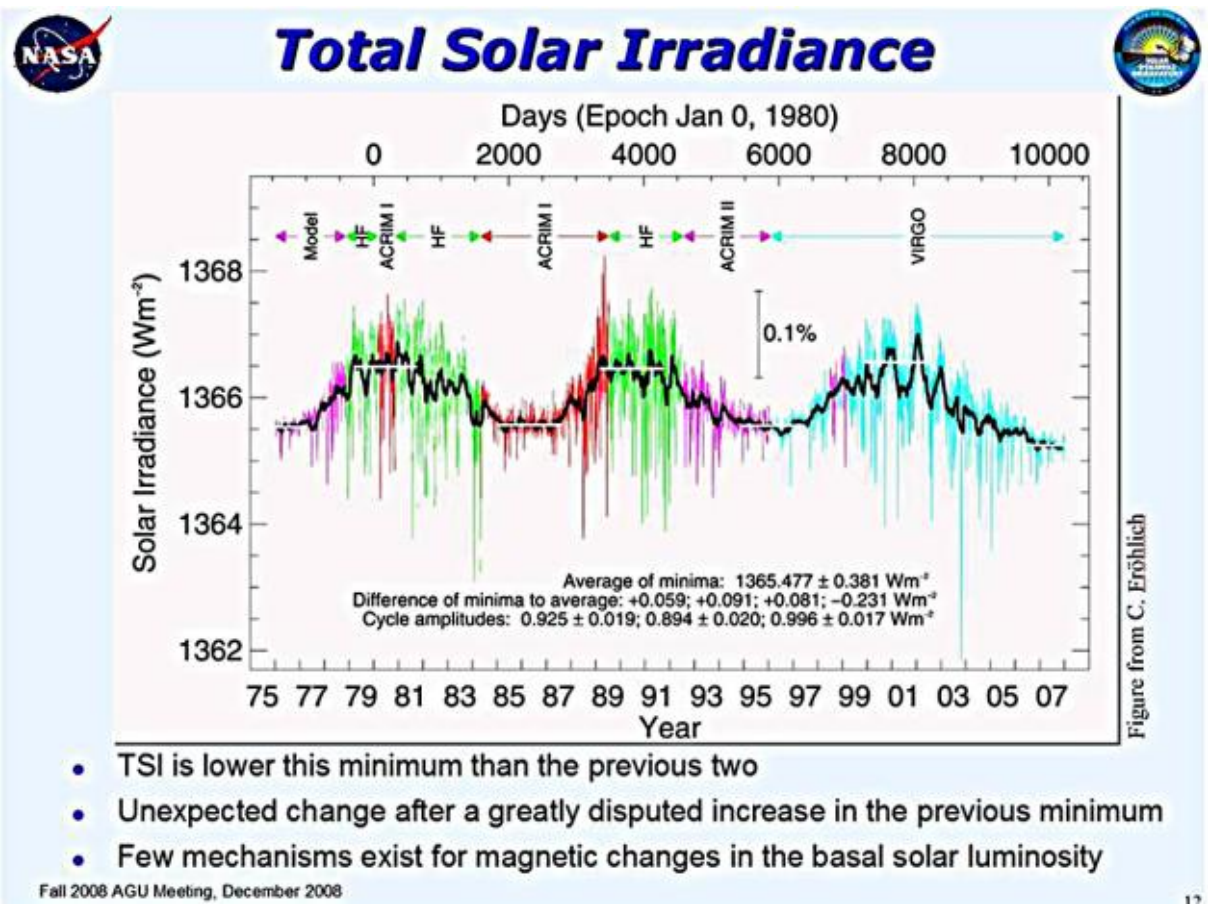

Figure 6- Variations of the TSI as measured during 3 solar cycles using different space-borne experiments.

Concerning $\delta T_{eff} / T_{eff}$ , measurements performed during 35 Years at Kitt Peak Observatory using the most powerful methods available today suggest no significant variations of the effective temperature of the photosphere in phase, or not, with the solar sunspot cycle, at the level of a noise which is still not sufficient to completely exclude any influence on the TSI [5]. We are then faced with the question of a possible dominant effect due to solar diameter variations, even if the theoretical aspects of the question seem not to be in favor of such explanation as the central nuclear source of solar energy is not believe to change over the human scales. But little details are known concerning the long processes bringing this energy to the surface of the Sun and possibly interacting with the solar activity cycle. To quantitatively translate solar magnetic modulation into irradiance variations, a clear mechanism-understanding is needed, or, better, evidences from solar "global" data are needed, including the fine analysis of solar radius variations. Note that solar radius measurements (see after) are performed since a long time (in the famous Secchi book of 1872 it was already discussed!) and a canonic value of 959"63 is still today almost universally adopted after the extended 40 Years long series of observations summarized and discussed by Auwers [6] where a correction of -1"55 for the irradiation effects was introduced. Needless to say that this is an arbitrary value coming from the discussion of visual type observations which is difficult to justify with today measurements with the use of CCD or CMos detectors and computer assisted type measurements. Another important aspect of the question is related to the definition to be adopted for the edge of the Sun used to measure the diameter. We will see that it is a difficult and controversial question which needs a special attention [8].

### *2.2- Different methods used to measure the solar diameter.*

After the long series of measurements of the 17th and of the 18th centuries using micrometers and transit timing, mainly by the French observers after Picard at the Paris Observatory, even more serious series were performed by British and German observers of the 19th century, based on the use of a specially designed instrument called the ***heliometer*** with an entrance aperture which could reach 20 cm. The instrument produced a dual image of the Sun and the solar diameter was deduced from the visual evaluation of their contact by measuring the angle of separation of the optical system. Visual evaluations are subject to the irradiation effect which means that a correction has to be introduced in addition to both instrumental and atmospheric effects [8]. A modern variant of this method has been introduced 30 Years ago based on the use of an adaptation of the Danjon astrolabe that is now called the solar astrolabe. A lot of efforts were made to improve the methods, including the use of computer assisted measurements and the use of a CCD detector but rather contradictory results were published and it is not possible here to discuss the details and make an evaluation of this new results. It is just worth mentioning that these new instruments use a really too small aperture to produce an image of the Sun and, accordingly, they are very sensitive to the Earth turbulence effects.

Another method takes advantage of the transit of the planet Mercury and even of Venus over the solar disk [7]. However such transits are rare and they are also subject to spurious effects due to the large amount of scattered light from the bright solar disk. Lately a Mercury transit was observed from Space using several imaging devices in EUV and interesting results were reported concerning the level of scattered light in the instrument. Another possibility to study the solar diameter variations free of atmospheric effects was offered by using the guiding telescope (operated in the visible light) of the rapidly spinning satellite used with the X-rays solar telescope RHESSI always pointed to the Sun. Long range relative variations due to the solar oblateness could be precisely analyzed but spurious effects due to the instrumental scattered light are also present and no solar diameter measurement can reasonably be made.

Another method which is worth mentioning is coming from the interpretation of heliosismic data. The frequencies of surface gravity waves, the so called f-modes, are related to the solar diameter and their precise determination could give a value of $R_0$. Unfortunately many different factors should also be taken into account and their influence is disputed [9].

The most precise measurements made until now from space take advantage of the Michelson Doppler Imager (MDI) telescope on board the Mission SoHO of ESA and NASA operated for more than 15 Years. Although the telescope was not designed for making solar diameter measurements a great number of images of the solar disk were made available permitting a deep analysis of the solar diameter variations. Variations were mainly due to instrumental effects (mainly thermo-mechanical stresses) and no significant truly solar variations has been found [10] above 0"023 peak to peak.

There again the amount of scattered light from the instrument does not permit a precise evaluation of the edge of the Sun due to the large illumination of the telescope by the light coming from the whole solar disk. Even the specially designed SODISM instrument of the Picard mission suffers from the same effect.

### *2.3- Solar eclipses to measure the solar diameter.*

Only at the time of a total eclipse of the Sun, the Moon occulting the full disk of the Sun in space (at a 400 000 km distance!), images made at the ground will be free of parasitic light. This occurs just before the full occultation, the true edge of the Sun being seen without the usual spurious amount of scattered light typical of non eclipse observations. Both the Earth atmosphere along the line of sight and the imaging instrument (or the photometer producing a light curve) are in the shadow of the Moon and they will not produce any spurious scattered light. It is one of the fundamental advantage of the eclipse method to measure the solar diameter and see the true edge of the Sun (it is also the reason why the solar corona is suddenly revealed during the total eclipse). Its precise angular value can be deduced from the timing of the contacts, see Fig. 7, by comparing the determined value at the eclipse with the ephemeride value predicted for the contacts and calculated using a standard value of the solar diameter.

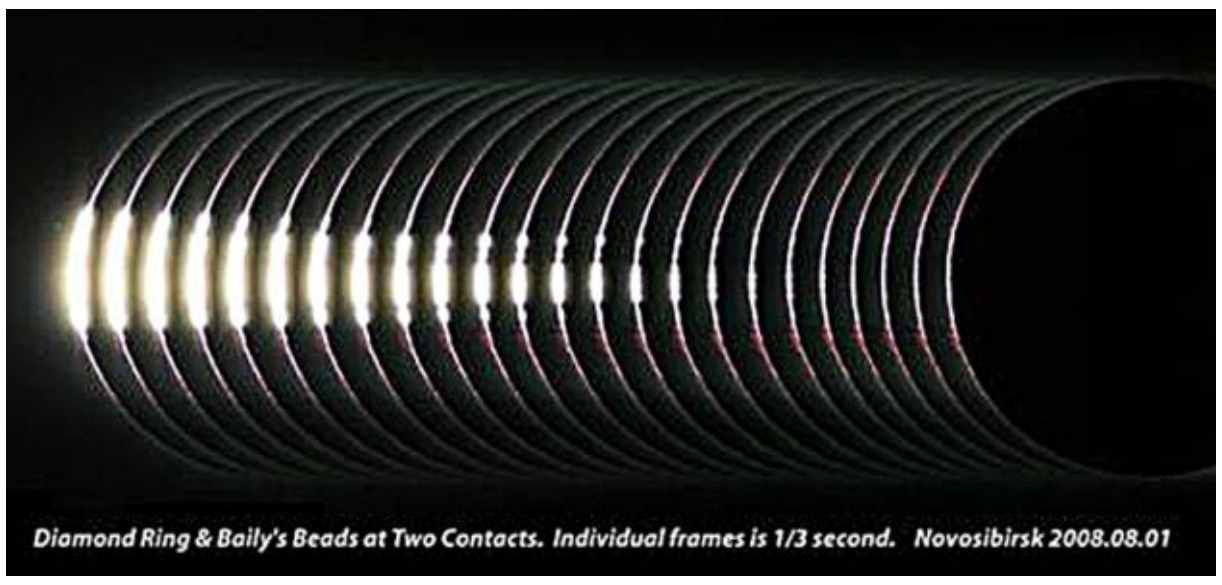

Figure 7- Images taken near the time of the 2d contact of a total eclipse of the Sun. Note the Baily beads due to the details of the Lunar limb.

The method is specially suitable for looking at long time range possible variations of the diameter of the Sun because the Moon is used as a reference and obviously the Moon diameter does not change in time at human scale. It is also a differential method as the occultation of the Sun occurs with a relative timing which is typically 30 times slower than the transit time of a celestial body in the sky due to the Earth rotation.

This method has been used for a long time [7] and even historical eclipses were considered for discussing the possible variations of the solar diameters [11]. Unfortunately, the method is also subject to errors due to effects produced by the changing details of the lunar limb for different libration angles. This effect produces the so called Baily beads, see Fig. 7 and the underestimation of their effect probably explains several claims made in the past decades of a large (typically 0”3 to 0”4) variation of the solar radius. Today much more precise measurements can be done using recent technologies including the timing and the site positioning with a GPS basis, the use of fast and precise photometers with extended dynamics and the use of large CMos imagers operating at a high cadence. Even more important, it is possible to record at the same cadence, flash spectra with an excellent dispersion and free of parasitic scattered light in order to look at the true edge of the Sun. Finally, it is also possible to improve the solar eclipse method by introducing more precise lunar profiles coming from the recently flown lunar altimetry experiments (for ex. the Kaguya space mission) to correct the former Watts profiles of the Moon limbs for different libration angles.

## 3. The 2010 solar eclipse experiments and preliminary results.

At the time of the last solar total eclipse of July 11, 2010 we prepared a set of experiments in order to perform eclipse measurements of the solar diameter. By chance, the totality occurring above the South Pacific Ocean could also be observed from several well separated atolls of the French Polynesia, see Fig. 8.

The program of observations included the study of the solar corona using white-light images and also a spectroscopic analysis during the inner contacts and during the totality, see Fig. 9.

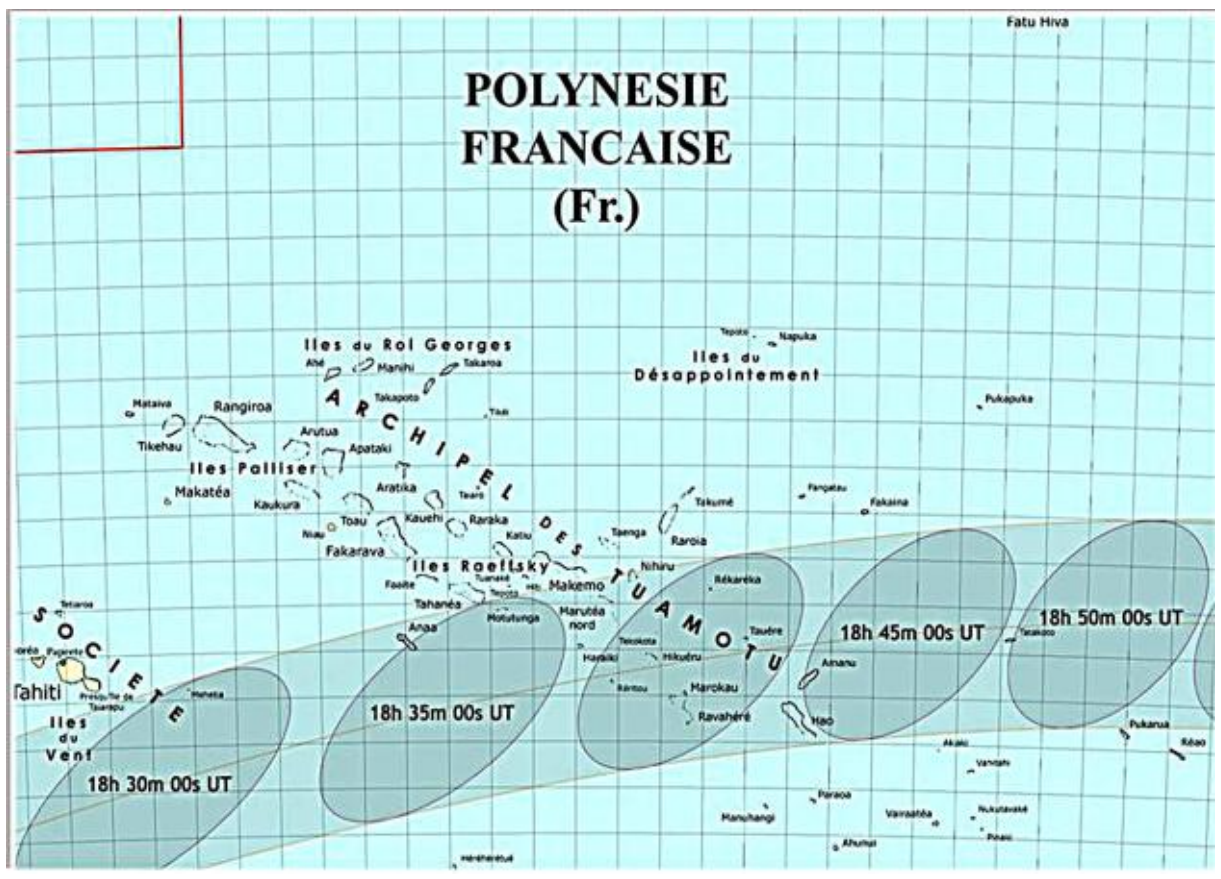

Figure 8- Map to show where the total eclipse was observable. Note the motion of the lunar shadow (map provided by IMCCE- Paris).

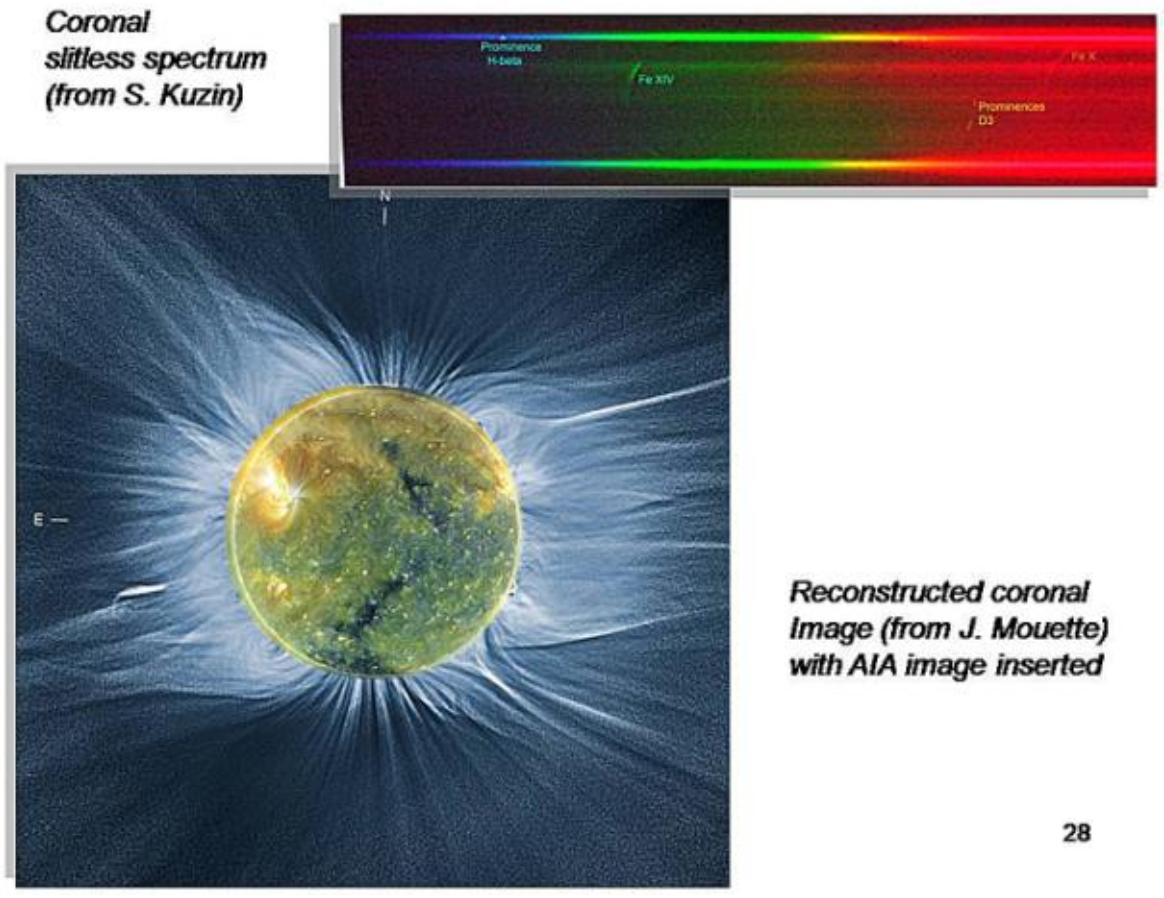

Figure 9- Eclipse composite image of the solar corona obtained at 18:43 UT with an EUV solar disk image from AIA (SDO) inserted in place of the image of the Moon. It is taken also during the total eclipse of the Sun but outside the shadow of the Moon. The corona in white light was observed by Jean Mouette of the Institut d’Astrophysique de Paris, CNRS and UPMC on the atoll of Hao, French Polynesia. At top right, a sample of slitless spectra obtained during the totality by Serguei Kuzin, from FIAN in Moscow (Russia).

### *3.1- Eclipse photometers to measure the solar diameter.*

Special photometers were designed by CNES (France) to make precise measurements of the contacts from several selected sites, including sites at the limit of the totality. This rather large experiment will be described in a forthcoming article and here we just mention some preliminary features. 12 photometers were put on different locations of several Atolls the day before the total eclipse and they were collected the day after. Because each photometer included the use of a GPS component, they could operate autonomously and provide an absolute timing of the contact. Some very small drift effect was however influencing the recordings and the effect is still evaluated. The dynamical range of measurements was large, many decades, as the measurements started just before the 1$^{st}$ contact and ended just after the 4$^{th}$ contact, more than 3 hours after. Unfortunately, clouds interfered with almost all the measurements but this did not really introduce a great uncertainty in the results concerning the timing of the contacts. The location of each photometer, including the altitude above the sea level, was determined with a great precision thanks to the GPS.

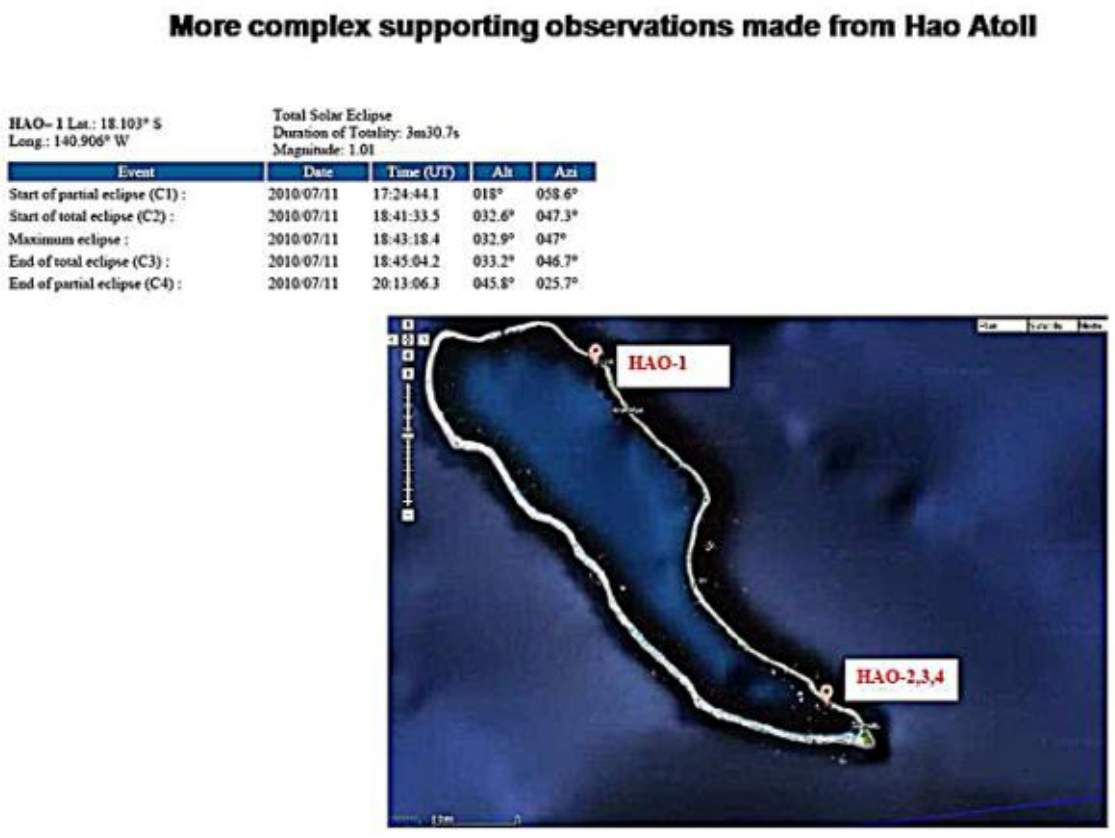

Figure 10- A map of Atoll Hao to show the locations of the photometers put to observe the contacts at different point of the lunar limbs.

To determine the best value of $R_0$ from the comparison of the measured contact time and the ephemerides, the most uncertain factor is the position and the details of the lunar limb, better defined using the mission Kaguya profiles which are not yet available with an absolute reference frame of the optical lunar limb. It is now the subject of a detailed study that is beyond the scope of this paper.

### *3.2- Eclipse observations and results from Hao.*

We now show some preliminary results coming from the main eclipse site on the atoll Hao, see Fig. 10.
Our team prepared, beside the eclipse photometers experiment see Fig. 11, several imaging and spectroscopic eclipse experiments.

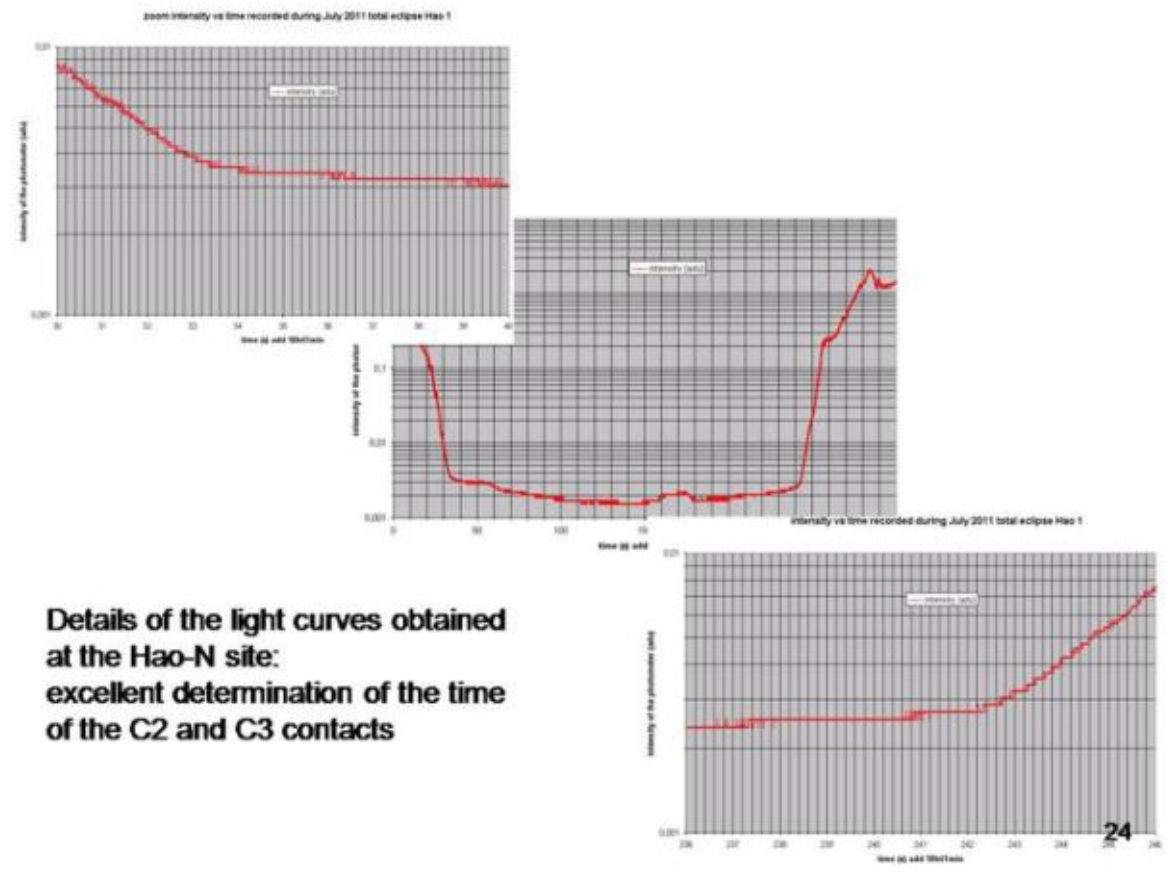

Figure 11- Sample of results from the recordings performed by the eclipse photometer put at the site Hao.

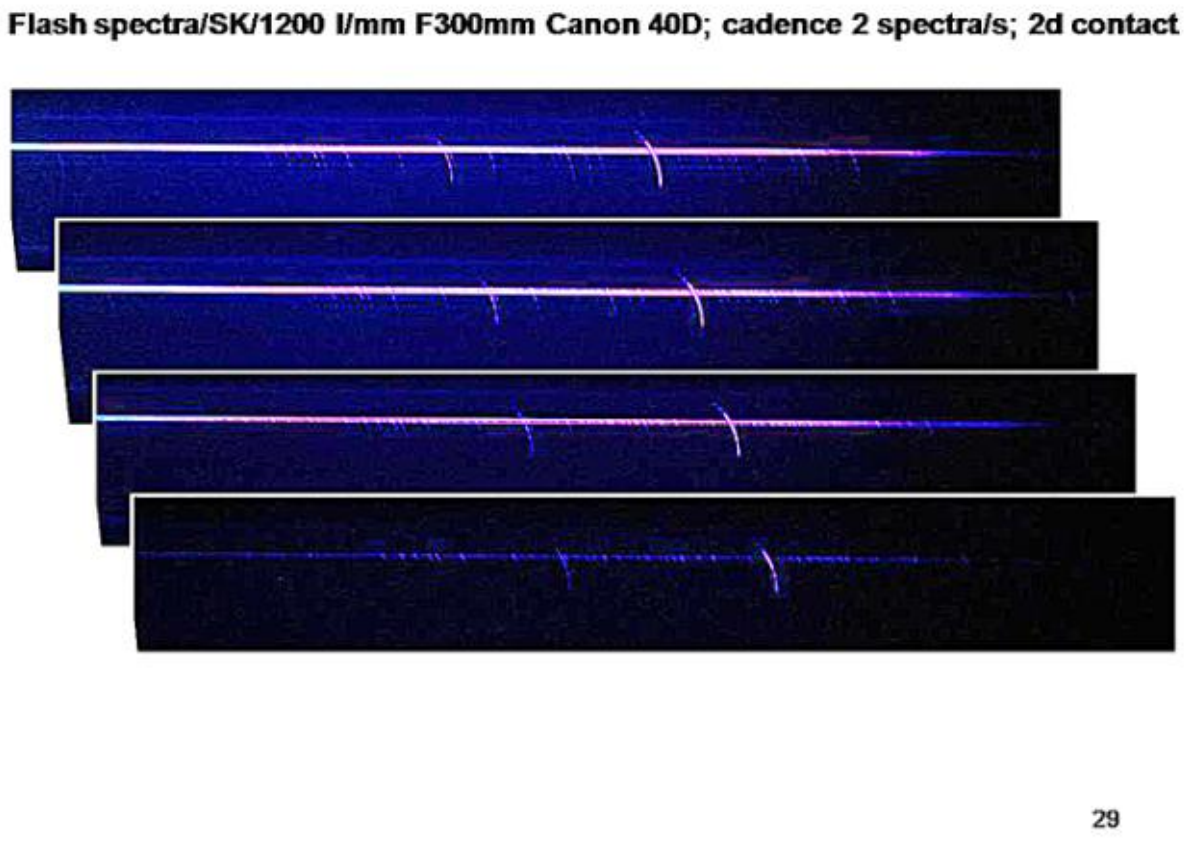

Figure 12- Poster made of several flash slitless spectra obtained during the 2d contact of the eclipse. The spectral range coverage goes from the deep blue to the green.

Fig. 12 illustrates the result from one of the spectroscopic experiment shooting color flash slitless spectra over a large spectral interval, at a cadence of 2 spectra per sec, in order to cover the very edge of the Sun. After, some longer exposure spectra were made to study the coronal lines produced above the solar limb when the chromospheres is fully covered by the Moon disk, see Fig. 9 the image of the corona in white light.

Fig. 12 already shows that the very edge of the solar disk, which is almost completely occulted by the irregular edge of the Moon, is indeed difficult to define. Although some continuum radiation with F- lines imprinted (seen in absorption) is still well recorded at a location where a valley of the Moon is showing the very bright edge of the Sun, a myriad of new low excitation emission lines is appearing just during a few seconds superposed to the continuum spectrum. Note that 1 sec of observation approximately corresponds to 300 km on the Sun. Outside an eclipse these lines, not to be confounded with the much more extended towards the corona hotter chromospheric lines, are not seen because the strong effect at the solar limb produced by the parasitic light coming from the bright disk. Indeed one of us (Cyril Bazin) developed a special technique for making fast flash slitless spectra during the contact and Fig. 13 illustrates the resolution achieved and shows some identification of the very faint emission lines superposed on the continuum spectrum of the very limb, indeed the true edge of the Sun.

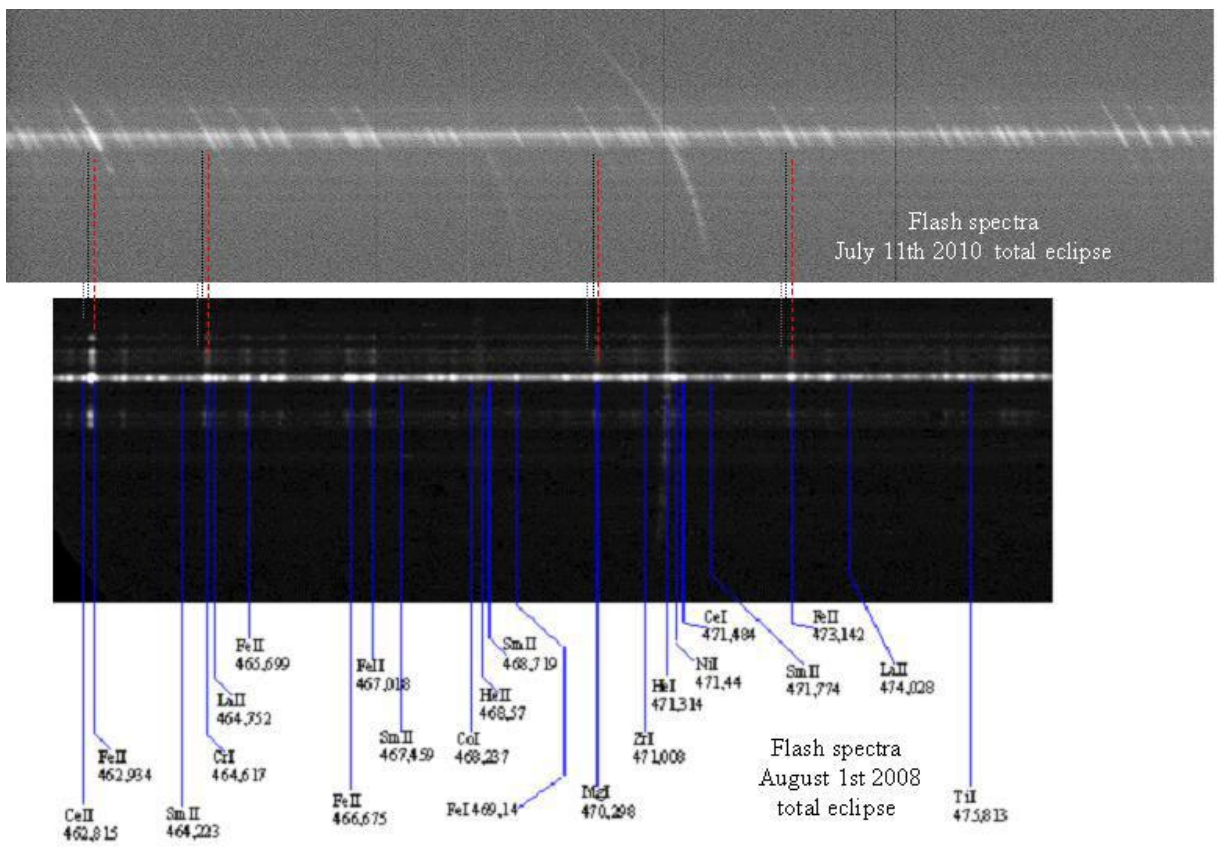

Figure 13- Sample of spectra obtained by Cyril Bazin using a fast CCD to analyze the flash slitless spectra during the last solar total eclipse. Some identification of the superposed faint emission lines on the very edge of the solar disk is shown at the bottom after making a linearization of the curved spectrum.

### *3.3- Discussion of the main Eclipse observations.*

Such result as shown on Fig. 12 and 13, coming from fast spectroscopic eclipse observations with the aim of observing the solar extreme-limb and eventually measure a solar diameter was a surprise. It needs a careful evaluation and more importantly, it is shedding some light on the past observations and measurements of the solar diameter because such measurements were never done with a spectral resolution good enough to resolve the faint low excitation lines appearing at the solar limb. Depending of the method used in the past (visual; photographic; CCD or CMos detector with filter or not), the influence of these lines could be different giving different values for the solar diameter, although the irradiation correction artificially introduced to correct the measured values includes a correction for this effect. Such correction is not needed for eclipse observations like the measurements done with the CNES eclipse photometers but the effect of the faint emission lines will have to be considered. However the continuum radiation radial variations supposed to define the limb of the Sun was indeed never correctly measured. It is also pointing to the need to have a better definition of the edge of the Sun, a problem which is now starting to be seriously considered [12] although no account for the formation of the faint emission lines is introduced because the used models are 1D and they make the assumption of hydrostatic equilibrium. The upper layers of the solar atmosphere are clearly dominated by the emergence of the magnetic field and this effect can be taken into account only when using a 3D simulation and a full set of MHD equations.

## 4. Conclusion

Measurements of the solar diameter present a great interest in the frame of the debate concerning the so-called solar forcing effect on the Earth climate. It is also a fundamental “constant” of astrophysical interest and finally it is related to the knowledge of the solar atmosphere and of magnetic phenomena occurring at the heights corresponding to the very extreme limb.

We presented a short review of the past observations and measurements with the feeling that they mainly represent a historical interest but that they cannot be taken seriously at the light of modern researches. Eclipse observations permit to more correctly evaluate the very edge of the Sun, see Fig. 12 and 13, but to provide valuable measurements of the solar diameter we need to know not only the details of the lunar limb but also the position of the lunar limb in an absolute reference optical frame, not just with respect to the centre of gravity of the Moon. This is explained in Fig. 14 using the former Watts lunar profiles where already some good correlations were observed with the observed spectra. New more detailed profiles are now available from the mission Kaguya and they will be used as soon as the profiles will be referenced in an optical system.

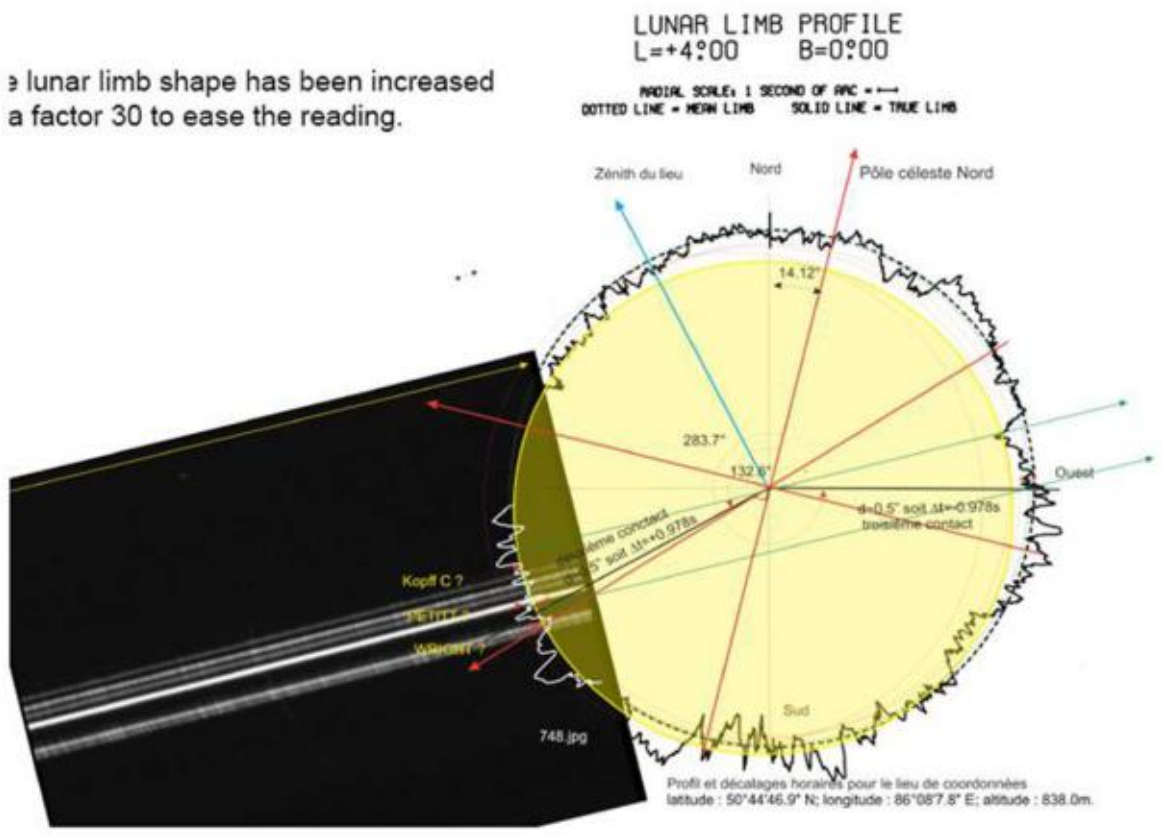

Figure 14- Schema to illustrate the method used to measure the solar diameter. The precise timing of the flash spectra should be referred to an optical position of the lunar edge in order to compute the angular diameter of the Sun.

The influence of the faint low excitation emission lines will have to be taken into account, after a consensus will be reached regarding the definition of the solar limb. This is a critical point in case of non eclipse observations when the extreme limb of the Sun is not properly measured because a large amount of scattered light is present. This means that special techniques like performing artificial eclipses in Space, could be used in the future. Natural eclipse observations will be improved in the future to provide measurements at long temporal range. Finally, the atmospheric models used to interpret the solar diameter variations should include the discussion of the influence of the faint emission lines and some additional observational works using modern CCD observations will have to be done to provide an atlas describing this effect.

**Acknowledgments**

It is a pleasure to thank Prof. Lotfia Nadi, from the Cairo University, for organizing a very successful high level international MTPR meeting and for inviting me to give an inaugural presentation at the Cairo University. The topic on the “discussion of the solar diameter measurements and the solar variabilities” was chosen after discussing with Prof. Ahmed Hady of the same University and we thank him for his help. Eclipse observations reported here are the result of efforts produced by many people who were involved at different stages of preparation and observations. It is difficult to list everybody but we want to name Jean Mouette, Patrick Martinez, Meleana Adams, Michel Lamiroté and Roland Santalo who directly participated in the observations reported here. Costantino Sigismondi also provided very helpful and valuable discussions during the development of this project.